\documentclass[12pt]{article}
\usepackage{amsfonts,amssymb,amsmath}
\usepackage{color}
\usepackage{graphicx,cite,epsfig,rotating}
\usepackage{pstricks}
\usepackage{slashed}
\usepackage[normalem]{ulem}

\advance\textheight by \topskip
 \newcommand{ \slashchar }[1]{\setbox0=\hbox{$#1$}   
    \dimen0=\wd0                                     
    \setbox1=\hbox{/} \dimen1=\wd1                   
    \ifdim\dimen0>\dimen1                            
       \rlap{\hbox to \dimen0{\hfil/\hfil}}          
       #1                                            
    \else                                            
       \rlap{\hbox to \dimen1{\hfil$#1$\hfil}}       
       /                                             
    \fi}                                             %
\begin{document}

\begin{center}
{\Large\bf Neutrino masses along with fermion mass hierarchy }\\[15mm]
{\large Dilip Kumar Ghosh\footnote{dilipghoshjal@gmail.com}
and R.~S.~Hundi\footnote{tprsh@iacs.res.in}}\\[4mm]
{\em Department of Theoretical Physics, \\ 
     Indian Association for the 
     Cultivation of Science, \\ 2A \& 2B Raja S. C. Mullick Road, Kolkata 700032, India}\\[10mm]
\end{center}
\begin{abstract}
Recently a new mechanism has been proposed to cure the 
problem of fermion mass hierarchy in the Standard Model (SM) model. In this
scenario, all SM charged fermions other 
than top quark arise from higher dimensional operators 
involving the SM Higgs field. This model also predicted
some interesting phenomenology of the Higgs boson. We 
generalize this model to accommodate neutrino masses ( Dirac \& Majorana )
and also obtain the mixing pattern in the leptonic sector. To generate
neutrino masses, we add extra three right handed neutrinos $(N_{iR})$ 
in this model.  

\end{abstract}


\newpage
\section{Introduction}

The experimental observations from several neutrino oscillation data 
indicate that neutrinos have mass of the order of ${\cal O}(10^{-10})$ GeV 
and they also mix 
(see \cite{Strumia:2006db} and \cite{GonzalezGarcia:2007ib} for review). 
In the Standard Model (SM), all three flavored neutrinos are left handed and 
massless. Hence, to generate massive neutrinos, one needs to invoke physics 
beyond the SM. Since neutrinos are electrically neutral, it can be either
Dirac or Majorana fermions. To generate the Dirac mass term for 
the neutrinos, one has to add right handed neutrinos in the particle contents
of the SM. On the other hand for Majorana neutrinos, one has to break the
lepton number which is an accidental symmetry of the SM.  

The seesaw mechanism \cite{seesaw1} has been identified as the most natural 
scenario to generate small masses for neutrinos. In this scenario, one adds
a heavy particle of mass $M$ in the SM, which after being integrated out, 
leads to the gauge invariant $D=5$ operator with only SM fields, 
${\cal L}_{eff} = y \frac{LLHH}{M} $, with $M >> M_W $ assumed. There 
are mainly three types of the seesaw mechanisms have been realized
depending upon the type of the exchanged heavy particles:
\begin{itemize} 
\item Type -I : Heavy right handed neutrinos are exchanged \cite{seesaw1, 
Mohapatra:1979ia}
\item Type -II : Heavy $SU(2)$ triplet scalars are exchanged \cite{seesaw2}
\item Type -III : Heavy $SU(2)$ neutral triplet fermions are exchanged 
\cite{Foot:1988aq}
\end{itemize} 

All the above mechanisms have been proposed to generate neutrino masses of 
the order of ${\cal O}(10^{-10})$ GeV, which immediately suggest that the
neutrinos are much lighter than their charged $SU(2)$ partners. Although,
we still do not know the exact masses of the neutrinos, but their mass
differences point out of some hierarchies among different generations,
which are very different from that of the charged leptons. In addition,
the observed neutrino mixing angles indicate strong flavor mixing in the
leptonic sector compared to the quark sector. As a result of this, the 
neutrino mass models are expected to explain not only the smallness of 
the masses but also the flavor structure of the lepton sector.

It has been be shown in Ref.\cite{babu_nandi} that by using higher dimensional
operators involving the relevant SM fermion fields 
and successive powers of the Higgs doublet field one could obtain a good fit to 
quark and charged lepton masses and mixing angles. These Yukawa interactions
can be expressed in powers of ${H^\dagger H}/{M^2}$, where $H$ is the
SM Higgs doublet field and $M$ is a mass scale $\sim {\cal O}(1-2)$ TeV 
at which SM would be embedded in an ultra violet (UV) theory. 
The dimensionless 
coefficients of these non-renormalizable operators which has 
inverse mass-dimensions can all be order one, which in turn leads to a
small Yukawa couplings of the SM in a natural way. This model also 
predicts some interesting Higgs phenomenology, for example, 
the enhanced $b {\bar b} H^0$, and $ \mu {\bar \mu} H^0$ couplings,
flavor changing Higgs boson decay $H \to {\bar t} c$ \cite{babu_nandi}. 
The strength of the flavor-changing ${\bar t}cH^0$ vertex is 
similar in magnitude of the flavor-conserving ${\bar b}b H^0$ vertex. 
However, this feature is not visible in the leptonic sector, as 
the flavor-changing $ \tau \mu H^0 $ vertex turns out to be two orders 
of magnitude suppressed compared to the flavor-conserving $\tau \tau H^0$ 
vertex. All these lead to a very interesting phenomenology which can 
be tested at the ongoing Large Hadron Collider (LHC) experiment. 

In Ref.\cite{babu_nandi}, authors have not addressed the issues of 
neutrino masses. In this paper, we try to obtain the right 
order of neutrino masses and the corresponding
Pontecorvo-Maki-Nakagawa-Sakata (PMNS) \cite{pmns1,pmns2} matrix by
introducing three additional right-handed neutrino fields $(N_{iR})$
in the above mentioned model. With these right-handed neutrinos, one 
can generate both Dirac and Majorana type neutrinos. We then compute the
PMNS matrix in both these two cases. 

Here, we would like to mention that generation of neutrino masses 
using effective operators higher than $d > 5 $ have been also 
discussed in Ref.\cite {Chen, Okada,Giudice, Babu_Nandi2,wwinter,Picek:2009is}. 
Most of these models have discussed only neutrino masses, but did
not attempt charged fermion masses simultaneously. Moreover, the mixing
pattern in lepton sector has also not been addressed in these
models. Some of these models have effective $d=6,7$ operators
with a TeV scale cutoff to explain the neutrino masses, but at the
expense of some more suppression through other couplings.
In contrast to the above mentioned models, we are trying to explain neutrino
masses and mixing pattern in a model which already has a natural mechanism
to explain the charged fermion masses. The effective operators we
consider have mass dimensions of 20 or 12, depending on Dirac or Majorana
neutrinos, and hence even with a TeV scale cutoff of the model we
do not have suppressions in dimensionless couplings. Since the neutrino
mass generation in this model correlates with that of charged fermion masses,
the parameters of this model are highly constrained, leading to
unique predictions for the phenomenology of this model.

The rest of the paper is organized as follows. In Sec.II, we briefly
discuss the model of Ref.\cite{babu_nandi}. In Sec.III we discuss the
mechanism of Dirac and Majorana neutrino mass generation.
In Sec.IV we discuss the
UV completion of the model. In Sec.V we outline some interesting 
phenomenology of this scenario. Finally, in Sec.VI we summarize our results. 

\section{The Model}
In the SM, the top quark mass, whose mass is around the electroweak
scale, can be explained naturally by the
corresponding Yukawa term. Whereas, the masses for other fermions
need suppressions in the respective Yukawa couplings. So in the
SM, the hierarchy in mass pattern of fermions reflects into
hierarchy in the corresponding Yukawa couplings. To address
this fermion mass hierarchy problem a model has been proposed
in \cite{babu_nandi}, where the suppression in Yukawa couplings, other
than the top quark, is explained through higher dimensional
terms. These higher dimensional terms are expressed
in powers of $\frac{H^\dagger H}{M^2}$, where $H$ is the
Higgs doublet of the SM and $M$ is a mass scale at which SM
would be embedded in a UV theory. The effective
terms in the SM to explain the fermion mass hierarchy
are written as \cite{babu_nandi}
\begin{eqnarray}
{\cal L}^{\rm Yuk} &=& h_{33}^u\bar{q}_{3L}u_{3R}\tilde{H}+
\left(\frac{H^\dagger H}{M^2}\right)(h_{33}^d\bar{q}_{3L}d_{3R}H
+h_{22}^u\bar{q}_{2L}u_{2R}\tilde{H}
+h_{23}^u\bar{q}_{2L}u_{3R}\tilde{H}+h_{32}^u\bar{q}_{3L}u_{2R}\tilde{H})
\nonumber \\
&& +\left(\frac{H^\dagger H}{M^2}\right)^2(h_{22}^d\bar{q}_{2L}d_{2R}H
+h_{23}^d\bar{q}_{2L}d_{3R}H 
+h_{32}^d\bar{q}_{3L}d_{2R}H+h_{12}^u\bar{q}_{1L}u_{2R}\tilde{H}+h_{21}^u\bar{q}_{2L}u_{1R}\tilde{H}
\nonumber \\
&&+h_{13}^u\bar{q}_{1L}u_{3R}\tilde{H}
+h_{31}^u\bar{q}_{3L}u_{1R}\tilde{H}) 
+ \left(\frac{H^\dagger H}{M^2}\right)^3
(h_{11}^u\bar{q}_{1L}u_{1R}\tilde{H}
+h_{11}^d\bar{q}_{1L}d_{1R}H \nonumber \\
&& +h_{12}^d\bar{q}_{1L}d_{2R}H
+h_{21}^d\bar{q}_{2L}d_{1R}H+h_{13}^d\bar{q}_{1L}d_{3R}H
+h_{31}^d\bar{q}_{3L}d_{1R}H)+{\rm h.c.}
\label{E:Ycop}
\end{eqnarray}
Here, $h^u$s and $h^d$s are ${\cal O}(1)$ couplings. Also, $q$s are left-handed
quark doublets, $u,d$ are singlet right-handed up- and down-type quark fields,
respectively. $\tilde{H}$ is the conjugate of $H$. The above higher
order terms can be explained from the UV completion of the SM, which
will be described later.

Terms in Eq. (\ref{E:Ycop}) are higher dimensional and generate
effective Yukawa couplings once 
the Higgs doublet acquires vacuum expectation value (vev). Although
the terms in Eq. (\ref{E:Ycop}) give masses to quarks, mass generation 
mechanism for
charged leptons is same as that for the down-type quarks. In the above
equation by replacing $q_{iL}\to L_{iL}$, $u_{iR}\to E_{iR}$ and 
$h^d_{ij}\to h^l_{ij}$, where $L$s and $E$s are left-handed doublet and 
right-handed singlet leptons, respectively, and $h^l$s are ${\cal O}(1)$ 
couplings, one would obtain mass terms for charged leptons. After the 
electroweak symmetry breaking the masses of quarks and charged
leptons will have a form \cite{babu_nandi}
\begin{eqnarray}
(m_t,m_c,m_u)&\approx &(|h_{33}^u|,|h_{22}^u|\epsilon^2,|h_{11}^u-h_{12}^uh_{21}^u/h_{22}^u|\epsilon^6)v,
\nonumber \\
(m_b,m_s,m_d)&\approx &(|h_{33}^d|\epsilon^2,|h_{22}^d|\epsilon^4,|h_{11}^d|\epsilon^6)v,
\nonumber \\
(m_\tau,m_\mu,m_e)&\approx &(|h_{33}^l|\epsilon^2,|h_{22}^l|\epsilon^4,|h_{11}^l|\epsilon^6)v,
\end{eqnarray}
where, $\epsilon=\frac{v}{M}$ and $v$ = 174 GeV is the vev
of the Higgs doublet. Along with the mass terms, we can
also get Cabbibo-Kobayashi-Maskawa (CKM) matrix in the quark sector.
It has been shown in \cite{babu_nandi} that
a good fit to the CKM matrix and to the masses of quarks and charged leptons can
be obtained for $\epsilon=1/6.5$, and the various ${\cal O}(1)$ couplings 
are found out to be\footnote{SM fermion masses are given in \cite{babu_nandi},
which include renormalization effects.}
\begin{eqnarray}
&&(|h_{33}^u|,|h_{22}^u|,|h_{11}^u-h_{12}^uh_{21}^u/h_{22}^u|)=(0.96,0.14,0.95),
\nonumber \\
&& (|h_{33}^d|,|h_{22}^d|,|h_{11}^d|)=(0.68,0.77,1.65),
\nonumber \\
&&(|h_{33}^l|,|h_{22}^l|,|h_{11}^l|)=(0.42,1.06,0.21).
\label{E:h_udl}
\end{eqnarray}
The value $\epsilon=1/6.5$ implies that $M\approx$ 1.1 TeV, which
is the scale at which a UV completion of the SM takes place.

The UV completion of this model is necessary in order to explain
the higher order terms of Eq.~(\ref{E:Ycop}). Consider a flavor symmetry
$G_F$ above the scale $M$, under which the third generation up-quarks
and Higgs boson are singlets and all other fermions transform
non-trivially. This charge assignment forbids the dimension-4
Yukawa terms for all fermions, expect for the top quark. Now, some
vector-like heavy fermions and complex scalar flavon fields $F$
with masses $\sim M$ can be proposed,
which transform under the flavor group $G_F$ but are singlets
under the SM gauge group. The role of these heavy vector-like
fermions and flavons $F$ is such that they form Yukawa-like terms with the SM
fermions at a high scale. The flavon fields $F$ can acquire vev around $M$ and
spontaneously break the flavor symmetry $G_F$. Upon integrating
the vector-like fermions, we
can generate higher dimensional terms of Eq. (\ref{E:Ycop}) \cite{babu_nandi},
where the dimensionless couplings $h^u$s and $h^d$s can be viewed
as functions of $\frac{\langle F\rangle}{M}$.

It is to be noted that the model in \cite{babu_nandi} can be generalized
by including an additional scalar singlet field $S$ \cite{LMN,BMN}. It
has been shown that instead of expanding in $\frac{H^\dagger H}{M^2}$,
the higher order terms of Eq. (\ref{E:Ycop}) can arise
in terms of $\frac{S^\dagger S}{M^2}$ \cite{LMN}. The model of this
kind is consistent and the UV completion of it has been
worked in detail \cite{LMN}. Likewise, we can also consider higher
order terms of Eq. (\ref{E:Ycop}) arising in expansion of both
$\frac{H^\dagger H}{M^2}$ and $\frac{S^\dagger S}{M^2}$ \cite{BMN}.
However, in this work, we stick to the minimal version of all these
models \cite{babu_nandi}, i.e. we do not assume extension to scalar Higgs
sector of the SM.

\section{Neutrino masses in this model}

In the above described model, neutrino masses have not been addressed, and
as a result we cannot also obtain the PMNS matrix in the
lepton sector. Here, we address both these issues by proposing
three right-handed neutrino fields ($N_{iR}$) into the model. However,
right-handed neutrinos can couple to left-handed neutrinos in such
a way that either Dirac or Majorana neutrinos can form. We study
both these cases in the following two subsections. 

\subsection{Dirac neutrinos}

From the neutrino oscillation data it is known that the atmospheric neutrino mass
scale ($\approx 0.05$ eV) is just a factor larger than the solar
neutrino mass scale ($\approx 0.009$ eV). This is to be compared to the
hierarchy of $\sim 10^4$ between electron and tau masses. Indeed, to
accommodate hierarchy between different family generations of charged fermions,
different powers of $\frac{H^\dagger H}{M^2}$ are assigned in Eq. (\ref{E:Ycop}),
which may not be necessary in the case of neutrinos because there are
no such large hierarchies within their masses.
Hence, from the above mentioned point and also from the naive order
of estimations on the neutrino mass scale, we propose the following
higher dimensional terms
\begin{equation}
{\cal L}^\nu_D=\left(\frac{H^\dagger H}{M^2}\right)^8h^\nu_{ij}\bar{L}_{iL}N_{jR}\tilde{H}.
\label{E:Dira}
\end{equation}
The mass dimension of the above operators are 20, which is large compared
to some dimension-10 operators in the quark sector of Eq. (\ref{E:Ycop}).
The largeness in the dimension for neutrino operators would give us very
small neutrino masses compared to the charged fermion masses.
The above higher order terms can be motivated by studying the UV completion
of this model, where we appropriately
choose the heavy vector-like fermions under the flavor group $G_F$ and
upon integrating them out we generate the above terms in the low energy regime.
The UV completion of this model will be described in the next section.
After the electroweak symmetry breaking, the above term gives Dirac masses for
neutrinos, which has a form
\begin{equation}
[M_D^\nu]_{ij}=\epsilon^{16}vh^\nu_{ij}.
\end{equation}

Now, our aim is to fit the atmospheric and solar neutrino mass-squared differences
and also the PMNS matrix with ${\cal O}(1)$ $h^\nu$ couplings. ${\cal O}(1)$ $h^\nu$
couplings mean that the values should be close to 1, but there is no well defined
range for these values. For example, in \cite{Giudice} ${\cal O}(1)$
couplings are meant to be in the range 1/5 to 5.
However, in this work we try for
$h^\nu$s to be between 0.1 and 2.0 because Yukawa couplings for charged fermions
are found to be within this range, see Eq. (\ref{E:h_udl}). When we present our
numerical results we will see that the $h^\nu$s may become slightly larger than
2.0 and we comment out over there. As for the PMNS matrix, it has some
specific structure and as result we would expect the couplings $h^\nu$ need to have some
structure as well. The PMNS matrix has been defined as $U_{\rm PMNS}=(V^l_L)^\dagger V^\nu_L$,
where $V^l_L$ and $V^\nu_L$ are unitary matrices which diagonalize the charged
lepton and neutrino mass matrices as follows:
\begin{eqnarray}
(V^l_L)^\dagger M^l(M^l)^\dagger V^l_L &=&{\rm diag}(m_e^2,m_\mu^2,m_\tau^2),
\nonumber \\
(V^\nu_L)^\dagger M_D^\nu(M_D^\nu)^\dagger V^\nu_L &=& {\rm diag}(m_1^2,m_2^2,m_3^2).
\end{eqnarray}
Here, $m_{1,2,3}$ are the three neutrino mass eigenvalues and $M^l$ is the
mass matrix in the charged lepton flavor basis, whose form is
\begin{equation}
M^l=\left(\begin{array}{ccc}
h_{11}^l\epsilon^6 & h_{12}^l\epsilon^6 & h_{13}^l\epsilon^6 \\
h_{21}^l\epsilon^6 & h_{22}^l\epsilon^4 & h_{23}^l\epsilon^4 \\
h_{31}^l\epsilon^6 & h_{32}^l\epsilon^4 & h_{33}^l\epsilon^2
\end{array}\right)v.
\end{equation}
We have found that the leading form of $V^l_L$ as
\begin{equation}
V^l_L=\left(\begin{array}{ccc}
1 & \frac{h^l_{12}}{h^l_{22}}\epsilon^2 & \frac{h^l_{13}}{h^l_{33}}\epsilon^4 \\
-\frac{h^l_{12}}{h^l_{22}}\epsilon^2 & 1 & \frac{h^l_{23}}{h^l_{33}}\epsilon^2 \\
-\frac{h^l_{13}h^l_{22}-h^l_{23}h^l_{12}}{h^l_{22}h^l_{33}}\epsilon^4 &
-\frac{h^l_{23}}{h^l_{33}}\epsilon^2 & 1
\end{array}\right).
\end{equation}
The above equation indicates that the form of $V^l_L$ is close to unit matrix
with the off-diagonal elements are suppressed by at least $\epsilon^2$. This
observation indicates that the unitary matrix $V^\nu_L$ should have nearly the
PMNS structure.

The PMNS matrix is determined by the three mixing angles and one CP violating
phase. In this work, for parameterization of PMNS matrix we have followed
the convention in \cite{pdg}. Before the data of T2K experiment, a global fit
to various neutrino oscillation data \cite{STV} gave results that the
$\theta_{13}$ was allowed to be zero at 2$\sigma$ level and the exact
tribimaximal mixing pattern \cite{HPS} in the lepton sector was still a
possibility. Recently, in the T2K experiment \cite{T2K} the
appearance of six events of electron-neutrinos in the detector has ruled out
$\theta_{13}\neq 0$ at 90 $\%$ C.L. However, the analysis of T2K is done
by putting $\theta_{12}\approx 34^{\rm o}$ and $\theta_{23}=45^{\rm o}$, which
suggests that the values of $\theta_{12}$ and $\theta_{23}$ are in agreement
with the corresponding tribimaximal values. To be consistent with the T2K experimental
result, we take the CP violating phase to be zero and the leptonic mixing angles
to be: $\sin\theta_{12}=\frac{1}{\sqrt{3}}$, $\sin\theta_{23}=\frac{1}{\sqrt{2}}$,
and $\sin\theta_{13}=0.157$. The $\sin\theta_{13}$ value gives $\theta_{13}\approx
9^{\rm o}$, which is consistent with the lower and upper bounds by the T2K \cite{T2K}
and CHOOZ experiments \cite{chooz}, respectively. We consider this
value for $\theta_{13}$ only to demonstrate that PMNS structure can be obtained
with ${\cal O}(1)$ Yukawa couplings, but otherwise it can be varied within the
experimental limits.

We take the unitary matrix $V^\nu_L$ as
\begin{equation}
V_L^\nu=\left(\begin{array}{ccc}
\sqrt{\frac{2}{3}}c_{13} & \frac{1}{\sqrt{3}}c_{13} & s_{13} \\
-\frac{1}{\sqrt{6}}-\frac{1}{\sqrt{3}}s_{13} & \frac{1}{\sqrt{3}}-\frac{1}{\sqrt{6}}s_{13}
& \frac{1}{\sqrt{2}}c_{13} \\
\frac{1}{\sqrt{6}}-\frac{1}{\sqrt{3}}s_{13} & -\frac{1}{\sqrt{3}}-\frac{1}{\sqrt{6}}s_{13}
& \frac{1}{\sqrt{2}}c_{13}
\end{array}\right),
\label{E:tribi}
\end{equation}
where $c_{13}=\cos\theta_{13},s_{13}=\sin\theta_{13}$. The above form of
$V^\nu_L$ is same as the PMNS matrix with mixing angles in the leptonic
sector, as mentioned above. Using the above $V^\nu_L$, we find
\begin{eqnarray}
(V^\nu_L)^\dagger M_D^\nu(M_D^\nu)^\dagger V^\nu_L &=& \epsilon^{32}v^2
\left(\begin{array}{ccc}
\vec{a}\cdot\vec{a} & \vec{a}\cdot\vec{b} & \vec{a}\cdot\vec{c} \\
\vec{a}\cdot\vec{b} & \vec{b}\cdot\vec{b} & \vec{b}\cdot\vec{c} \\
\vec{a}\cdot\vec{c} & \vec{b}\cdot\vec{c} & \vec{c}\cdot\vec{c}
\end{array}\right),
\end{eqnarray}
where the 3-dimensional vectors are: $\vec{a}=(a_1,a_2,a_3)$, $\vec{b}=(b_1,b_2,b_3)$ and
$\vec{c}=(c_1,c_2,c_3)$, with
\begin{eqnarray}
a_j&=&\sqrt{\frac{2}{3}}c_{13}h^\nu_{1j}-(\frac{1}{\sqrt{6}}+\frac{1}{\sqrt{3}}s_{13})
h^\nu_{2j}+(\frac{1}{\sqrt{6}}-\frac{1}{\sqrt{3}}s_{13})h^\nu_{3j},\quad
\nonumber \\
b_j&=&\frac{1}{\sqrt{3}}c_{13}h^\nu_{1j}+(\frac{1}{\sqrt{3}}-\frac{1}{\sqrt{6}}s_{13})
h^\nu_{2j}-(\frac{1}{\sqrt{3}}+\frac{1}{\sqrt{6}}s_{13})h^\nu_{3j},\quad
\nonumber \\
c_j&=&s_{13}h^\nu_{1j}+\frac{1}{\sqrt{2}}c_{13}(h^\nu_{2j}+h^\nu_{3j}).
\label{E:abc}
\end{eqnarray}
The necessary conditions to
be satisfied in order to fit the neutrino mass-square differences are
\begin{eqnarray}
&& m_1^2=f_D^2\vec{a}\cdot\vec{a},\quad m_2^2=f_D^2\vec{b}\cdot\vec{b}, \quad
m_3^2=f_D^2\vec{c}\cdot\vec{c},
\nonumber \\
&& \vec{a}\cdot\vec{b}=\vec{b}\cdot\vec{c}=\vec{a}\cdot\vec{c} =0,
\nonumber \\
&& m_3^2-m_1^2= \Delta m^2_{\rm atm}=(\pm 2.4)\times 10^{-3}~{\rm ev}^2,
\nonumber \\
&& m_2^2-m_1^2= \Delta m^2_\odot=7.6\times 10^{-5}~{\rm ev}^2,
\label{E:Dirsol}
\end{eqnarray}
where $f^2_D=\epsilon^{32}v^2$.

In Eq. (\ref{E:Dirsol}) we take the central values of the atmospheric and
solar neutrino mass-squared differences as they are considered in T2K 
experiment \cite{T2K}. Since the sign
of $\Delta m^2_{\rm atm}$ is not known in experiments, its value could
be either of the values given above.
We can take the form the vectors as $\vec{a}=|\vec{a}|(1,0,0)$, $\vec{b}=|\vec{b}|(0,1,0)$
and $\vec{c}=|\vec{c}|(0,0,1)$, to satisfy the orthogonality among these vectors. The
magnitude of these vectors ($|\vec{a}|,|\vec{b}|,|\vec{c}|$) can be determined
by fitting to the neutrino mass-squared differences.
If the sign of $\Delta m^2_{\rm atm}$ is positive, we get the normal hierarchical pattern
among the neutrino masses. In this case
we can take $m_1\sim \sqrt{\Delta m_\odot^2}$, $m_2=\sqrt{\Delta m_\odot^2+m_1^2}$,
and $m_3=\sqrt{\Delta m_{\rm atm}^2+m_1^2}$. Whereas,
in the case where the sign of $\Delta m^2_{\rm atm}$ is negative, the spectrum of neutrino mass
eigenstates is called inverted hierarchy, in which case
we can take $m_3\sim \sqrt{\Delta m_{\rm atm}^2}$, $m_1=\sqrt{\Delta m_{\rm atm}^2+m_3^2}$,
and $m_2=\sqrt{\Delta m_\odot^2+m_1^2}$. For a definite value of $\epsilon$, and
after finding the values of $|\vec{a}|$, $|\vec{b}|$ and $|\vec{c}|$, by
inverting Eq. (\ref{E:abc}), we compute the full neutrino Yukawa couplings. 

In the determination of $|\vec{a}|$, $|\vec{b}|$ and $|\vec{c}|$, the value of
$\epsilon$ should be fixed. We fix $\epsilon=1/6.5\approx 0.15$ as this value has given
a good fit to charged fermion masses and also the CKM matrix.
In the normal hierarchy, for $m_1=0$ and $m_1=0.7\times\sqrt{\Delta m^2_{ \odot}}$,
the Yukawa couplings, respectively, turns out to be
\begin{equation}
h^\nu=\left(\begin{array}{ccc}
0 & 0.29 & 0.45\\
0 & 0.26 & 1.99 \\
0 & -0.33 & 1.99
\end{array}\right),\quad
\left(\begin{array}{ccc}
0.29 & 0.35 & 0.45\\
-0.18 & 0.32 & 2.01 \\
0.11 & -0.40 & 2.01
\end{array}\right).
\end{equation}
In the above case, if $m_1$ is less than 0.7 times of $\sqrt{\Delta m^2_{ \odot}}$,
the element $h^\nu_{31}$ may become less than 0.1.
In the inverted hierarchy, for $m_3=0$ and $m_3=0.5\times\sqrt{\Delta m^2_{ \rm atm}}$,
the Yukawa couplings, respectively, come out to be
\begin{equation}
h^\nu=\left(\begin{array}{ccc}
2.31 & 1.66 & 0 \\
-1.43 & 1.49 & 0 \\
0.91 & -1.86 & 0
\end{array}\right),\quad
\left(\begin{array}{ccc}
2.58 & 1.84 & 0.22 \\
-1.59 & 1.66 & 0.99 \\
1.01 & -2.08 & 0.99
\end{array}\right).
\end{equation}
The above neutrino Yukawa couplings are ${\cal O}(1)$. As stated
before, compared to charged fermion Yukawa couplings whose values are
within 0.1 to 2.0, some elements of neutrino Yukawa couplings are
slightly above 2.0, especially in the inverted hierarchical case.
We have found that by increasing $\epsilon$ value to 0.16, the
neutrino Yukawa couplings would be in the range of 0.1 to 2.0 in both
the hierarchical cases. By increasing $\epsilon$ to 0.16, do not change
the charge fermion Yukawa couplings very much. In fact, we have
also checked that for $\epsilon = 0.16$ the CKM matrix elements
can be fitted to their experimental values. We will comment
more on the possible values of $\epsilon$ in Sec. 5, where we
describe upper limits on it arising due to $D^0-\bar{D}^0$ mixing.
Finally, we compute the PMNS matrix which is given by $(V^l_L)^\dagger V_L^\nu$.
This matrix depends
on the charged lepton Yukawa couplings, for which only the diagonal elements
have been computed from their mass relations. If we set all the off-diagonal
couplings for charged leptons
to be 0.5, the actual PMNS matrix in this model for $\epsilon= 1/6.5$ is
\begin{equation}
U_{\rm PMNS}=\left(\begin{array}{ccc}
0.81 & 0.56 & 0.15 \\
-0.50 & 0.54 & 0.68 \\
0.30 & -0.63 & 0.72
\end{array}\right).
\label{E:pmns}
\end{equation}
The above matrix elements are within the error limits
of the computed values through the mixing angles in neutrino
oscillation experiments. By changing the off-diagonal Yukawa couplings
of charged leptons to 1.5, we have found that the elements in
the above matrix will change only in the second decimals.

\subsection{Majorana neutrinos}

In the previous subsection we have analyzed the case of Dirac neutrinos
in the model \cite{babu_nandi}. In the case of Dirac neutrinos the fields
$N_{iR}$ are just the right-handed components of neutrinos and lepton
number is conserved. However, the phenomenology would change if the
lepton number is assumed to be violated, and we can have massive
sterile neutrinos in the model. Since the UV cutoff of the model is
$M\sim$TeV, we can choose the masses of the sterile neutrinos to be
slightly less than of the order of TeV. These TeV sterile neutrinos,
with some mixing with active neutrinos,
can give significant contribution to neutrinoless double beta decay
process and some collider processes as well.

Suppose that the following effective operators exist in the model.
\begin{equation}
{\cal L}_M = \left(\frac{H^\dagger H}{M^2}\right)^4h^\nu_{ij}\bar{L}_{iL}
N_{iR}\tilde{H} + \frac{M_i}{2}\overline{N^c_{iR}}N_{iR} + {\rm h.c.},
\label{E:Majo}
\end{equation}
where $M_i$, $i=1,2,3$, are the masses of the right-handed sterile neutrinos, and
the indices $i,j$ should be summed over 1,2,3. $N^c_{iR}$ is the charge
conjugate of $N_{iR}$. There are dimension-12 operators in the above
equation which can be 
motivated from the UV completion of this model, and
it will be presented in the next section. The masses of right-handed neutrinos
would be around TeV which is much larger than the Dirac neutrino masses
of the first term in the above equation. Hence, after integrating the
heavy right-handed neutrinos, the masses of light neutrinos are given by
\begin{equation}
(M_M^\nu)_{ij} = \epsilon^{16}v^2\sum_{k=1}^3 h^\nu_{ik}\frac{1}{M_k}h^\nu_{jk}.
\end{equation}
Since the above mass matrix is symmetric, it can be diagonalized by a
unitary matrix $V_L^\nu$ as 
\begin{equation}
(V_L^\nu)^{\rm T}M_M^\nu V_L^\nu={\rm diag}(m_1,m_2,m_3).
\end{equation}
Since we have argued previously that the corresponding unitary
matrix $V_L^l$ in the charged lepton sector is close to
unit matrix, we choose $V_L^\nu$ to have the form in Eq. (\ref{E:tribi}).
Then the above matrix
relation can be satisfied if the following relations are hold:
\begin{eqnarray}
&&m_1=f_M\vec{a^\prime}\cdot\vec{a^\prime},\quad m_2=f_M\vec{b^\prime}\cdot\vec{b^\prime},\quad
m_3=f_M\vec{c^\prime}\cdot\vec{c^\prime},
\nonumber \\
&&\vec{a^\prime}\cdot\vec{b^\prime} =\vec{a^\prime}\cdot\vec{c^\prime} =
\vec{b^\prime}\cdot\vec{c^\prime} =0,
\end{eqnarray}
where $f_M=\epsilon^{16}v^2$. The vectors $\vec{a^\prime} = (a_1^\prime,a_2^\prime,a_3^\prime)$,
$\vec{b^\prime} = (b_1^\prime,b_2^\prime,b_3^\prime)$
and $\vec{c^\prime} = (c_1^\prime,c_2^\prime,c_3^\prime)$ are such that
$a_i^\prime = \frac{a_i}{\sqrt{M_i}}$, $b_i^\prime = \frac{b_i}{\sqrt{M_i}}$ and
$c_i^\prime = \frac{c_i}{\sqrt{M_i}}$, where $a_i,b_i,c_i$ are defined
in Eq. (\ref{E:abc}). To satisfy the orthogonality in these vectors
we choose their forms as $\vec{a^\prime} =|\vec{a^\prime}|(1,0,0)$,
$\vec{b^\prime} =|\vec{b^\prime}|(0,1,0)$, $\vec{c^\prime} =|\vec{c^\prime}|(0,0,1)$. Like in the
previous Dirac neutrinos, in this case also to fit the neutrino mass-squared
differences we analyze both the normal and inverted hierarchical
mass spectrums of neutrinos. The mass eigenvalues of the three neutrinos
in terms of mass-squared differences are mentioned in the previous
subsection. After finding the $|\vec{a^\prime}|,|\vec{b^\prime}|,
|\vec{c^\prime}|$ and for definite values of right-handed neutrino masses,
we can invert the relations in Eq. (\ref{E:abc}) to find the Yukawa couplings.
The masses of right-handed neutrinos could be either degenerate or
non-degenerate, and so we study both these cases below. 
In our numerical analysis we have fixed $\epsilon=1/6.5$.
\\
\noindent
\\
{\bf Case I: Degenerate right-handed neutrinos}\\

We choose the degenerate mass scale of right-handed neutrinos to be
1 TeV, i.e. $M_1=M_2=M_3=$ 1 TeV. In the normal hierarchy,
for $m_1=0$ and $m_1=0.5\times \sqrt{\Delta m^2_\odot}$, the Yukawa couplings came
out to be, respectively, as
\begin{equation}
h^\nu=\left(\begin{array}{ccc}
0 & 0.98 & 0.64 \\
0 & 0.88 & 2.83 \\
0 & -1.10 & 2.83
\end{array}\right),\quad
\left(\begin{array}{ccc}
0.98 & 1.03 & 0.64 \\
-0.60 & 0.93 & 2.84 \\
0.38 & -1.16 & 2.84
\end{array}\right).
\end{equation}
In the inverted hierarchy, for $m_3=0$ and $m_3=0.5\times\sqrt{\Delta m^2_{\rm atm}}$,
the Yukawa couplings came out to be, respectively, as
\begin{equation}
h^\nu=\left(\begin{array}{ccc}
3.27 & 2.33 & 0 \\
-2.02 & 2.10 & 0 \\
1.29 & -2.62 & 0
\end{array}\right),\quad
\left(\begin{array}{ccc}
3.46 & 2.46 & 0.45 \\
-2.14 & 2.21 & 2.00 \\
1.36 & -2.77 & 2.00
\end{array}\right).
\end{equation}
Notice here that in the inverted hierarchical case, the element $h^\nu_{11}$ of neutrino
Yukawa coupling is even larger than 3.0. This situation can be improved
if we raise $\epsilon$ to 0.16, in which case $h^\nu_{11}$ would be between
2.0 and 3.0. Further increasing $\epsilon$ to 0.17 would bring all the neutrino
Yukawa couplings in the range of 0.1 to 2.0 in both the hierarchical cases.
However, $\epsilon=0.17$ may be ruled out by the $D^0-\bar{D}^0$ mixing
in this model, which will be described in Sec. 5.
\\
\noindent
\\
{\bf Case II: Non-degenerate right-handed neutrinos}\\

To illustrate how the neutrino Yukawa couplings change from
Case I, we take the following values for the three right-handed
neutrinos: $M_1$ = 500 GeV, $M_2$ = 800 GeV and $M_3$ = 1 TeV.
In the normal hierarchy, for $m_1=0$ and
$m_1=0.5\times \sqrt{\Delta m^2_\odot}$, the Yukawa couplings came
out to be, respectively, as
\begin{equation}
h^\nu=\left(\begin{array}{ccc}
0 & 0.87 & 0.64 \\
0 & 0.78 & 2.83 \\
0 & -0.98 & 2.83
\end{array}\right),\quad
\left(\begin{array}{ccc}
0.69 & 0.92 & 0.64 \\
-0.43 & 0.83 & 2.84 \\
0.27 & -1.04 & 2.84
\end{array}\right).
\end{equation}
In the inverted hierarchy, for $m_3=0$ and $m_3=0.5\times\sqrt{\Delta m^2_{\rm atm}}$,
the Yukawa couplings came out to be, respectively, as
\begin{equation}
h^\nu=\left(\begin{array}{ccc}
2.31 & 2.08 & 0 \\
-1.43 & 1.88 & 0 \\
0.91 & -2.34 & 0
\end{array}\right),\quad
\left(\begin{array}{ccc}
2.44 & 2.20 & 0.45 \\
-1.51 & 1.98 & 2.00 \\
0.96 & -2.47 & 2.00
\end{array}\right).
\end{equation}
Comparing the values of neutrino Yukawa couplings in this case with that
of Case I, we can notice that by decreasing the value of a right-handed
neutrino mass decreases the elements in the corresponding column of the
matrix $h^\nu$. For example, by decreasing the value of $M_1$ from 1 TeV
to 500 GeV, we can see that the elements in the first column of $h^\nu$
have decreased. This fact can be understood from the expressions of
$a^\prime_i,b^\prime_i,c^\prime_i$ and also the from the form of these
vectors that we have considered.

Finally, the PMNS matrix in the Majorana neutrino case is same as that
of Eq. (\ref{E:pmns}). The construction of PMNS matrix in this model
is such that it depends on $\epsilon$, charged lepton Yukawa couplings
and mixing angles which we have mentioned before. As a result of this,
either in the Case I or Case II
of Majorana neutrinos, we do get the same PMNS matrix as long as we
do not change the $\epsilon$ and charge lepton Yukawa couplings. This
suggests that in this model the right-handed neutrino masses do not
show up in the PMNS matrix, rather their implications are felt
in the neutrino Yukawa couplings.

\section{UV completion of the model}

In this section we describe how the higher order terms of Eqs.
(\ref{E:Dira}) and (\ref{E:Majo}) can be motivated from the UV
completion of the model. Our procedure of UV completion is
similar to that described in \cite{babu_nandi,LMN,BMN}. However,
here we do not fix the charges of any field and moreover we
study the higher order terms of any power in $\frac{H^\dagger H}{M^2}$.
In the case of Dirac and Majorana neutrinos of this model, we need
8 and 4 powers of $\frac{H^\dagger H}{M^2}$, respectively, in the
relevant higher dimensional operators. In our
approach, we first study how a single power of $\frac{H^\dagger H}{M^2}$
is possible in a higher dimensional operator of $\frac{H^\dagger H}{M^2}
\bar{L}_LN_R\tilde{H}$. For simplicity, here we have suppressed family
indices. Then in the next step, we try to see how two powers
of $\frac{H^\dagger H}{M^2}$ can arise in the $\left(\frac{H^\dagger H}{M^2}\right)^2
\bar{L}_LN_R\tilde{H}$. Then afterwards, we can generalize this procedure
to obtain any higher dimensional operator containing a finite
power of $\frac{H^\dagger H}{M^2}$. Since the origin of higher
order terms in quark and charged lepton sectors have already been
discussed in \cite{babu_nandi,LMN,BMN}, below we confine only
to the neutrino sector. However, we believe our procedure, with
a little modification, can be merged with that of \cite{babu_nandi,LMN,BMN}.
Otherwise, our procedure can be extended even to the quark and
charged lepton sectors.

As stated before, we have to propose a flavor symmetry group $G_F$
to forbid the leading Yukawa couplings in the neutrino sector. The
symmetry $G_F$ is an exact symmetry at and above the scale $M\sim$ TeV.
We take $G_F$ to be a gauged abelian symmetry, i.e. $G_F$ = U(1)$_F$. In a
first step to generate the higher order term $\frac{H^\dagger H}{M^2}
\bar{L}_LN_R\tilde{H}$, we propose vector-like, color-singlet fermionic
fields $K_1$ and $G_1$,
which are singlet and doublet, respectively, under the SU(2)$_L$ of
the SM gauge group. We also propose complex scalar flavon field
$F_1$, whose vev ($\langle F_1\rangle\sim M$)
spontaneously breaks the flavor symmetry U(1)$_F$, but otherwise
is singlet under the standard model gauge group. Now, consider the following
renormalizable terms at the high scale.
\begin{equation}
y_1\bar{L}_LK_{1R}\tilde{H} + y_2F_1\bar{K}_{1R}K_{1L} + y_3\bar{K}_{1L}G_{1R}H^\dagger
+ y_4F_1\bar{G}_{1R}G_{1L} + y_5\bar{G}_{1L}N_RH,
\label{E:rto1}
\end{equation}
where $y$s are ${\cal O}$(1) dimensionless couplings.

The above renormalizable terms can be justified by assigning
the following charges under U(1)$_F$: $L_L=K_{1R}=l,F_1 = f_1,
K_{1L}=G_{1R} = l-f_1,G_{1L}=N_R = l-2f_1,H = 0$. Here, the Higgs
doublet is uncharged under U(1)$_F$ and $l$ and $f_1$ are U(1)$_F$ charges
of lepton doublet and $F_1$, respectively. After spontaneous breaking
of U(1)$_F$ and after integrating out the heavy fermions ($K_1,G_1$), the above
terms generate $\frac{H^\dagger H}{M^2}\bar{L}_LN_R\tilde{H}$ in the
low energy regime. Notice here that by assigning different U(1)$_F$ charges
of the three lepton doublets and by proposing three different copies of $F_1$
field, i.e. $F_{1i},i=1,2,3$, the above procedure can be generalized to give
the full 3$\times$3 family structure in $\frac{H^\dagger H}{M^2}\bar{L}_LN_R\tilde{H}$.
In this generalization, we need three copies of $K_1,G_1$ and moreover
all the three right-handed neutrinos will have same U(1)$_F$ charge.

Let us note here that the field content proposed in Eq. (\ref{E:rto1}) somewhat resembles
to that of \cite{Lindner}, where a general study of seesaw Dirac neutrino masses has been studied with
arbitrary number of active and sterile neutrinos. Here the weak-singlet fields $N_R,K_{1L},K_{1R}$
are sterile and they have mixing masses with the active neutrino ($\nu_L$)
of $L_L$. The mixing between $K_{1L}$ and $N_R$ arises after integrating the heavy weak-doublet $G$.
As a result, we get 4$\times$4 mixing mass matrix among the above said fields. The
structure of this matrix is same as that of 4$\times$4 texture proposed in \cite{Lindner}.
In another context of these type of extended seesaw models, leptogeneis has also been studied \cite{exsesa}.

In the next step, to generate the higher order term $\left(\frac{H^\dagger H}{M^2}\right)^2
\bar{L}_LN_R\tilde{H}$, in addition to the above field content, we propose $K_2,G_2$ whose quantum numbers under
SM gauge group are same as that of $K_1,G_1$, and also a complex scalar
flavon field $F_2$, which is a singlet under the standard model gauge group.
Now, consider the following renormalizable terms at the high scale.
\begin{eqnarray}
&&y_1\bar{L}_LK_{1R}\tilde{H}+y_2F_1\bar{K}_{1R}K_{1L}+y_3\bar{K}_{1L}G_{1R}H^\dagger
+y_4F_2\bar{G}_{1R}G_{1L}+y_5\bar{G}_{1L}K_{2R}H
\nonumber \\
&&+y_6F_2\bar{K}_{2R}K_{2L}
+y_7\bar{K}_{2L}G_{2R}H^\dagger+y_8F_1\bar{G}_{2R}G_{2L}+y_9\bar{G}_{2L}N_RH.
\end{eqnarray}
Here, $y$s are ${\cal O}(1)$ dimensionless couplings.
The above terms can be justified with the following U(1)$_F$ charges:
$L_L=K_{1R}=l,F_1=f_1,K_{1L}=G_{1R}=l-f_1,F_2=f_2,G_{1L}=K_{2R}=l-f_1-f_2,
K_{2L}=G_{2R}=l-f_1-2f_2,G_{2L}=N_R=l-2f_1-2f_2,H=0$.
After spontaneous symmetry
breaking and also after integrating the heavy fields we can generate
$\left(\frac{H^\dagger H}{M^2}\right)^2\bar{L}_LN_R\tilde{H}$ in the low
energy regime. Again, to generate the full 3$\times$3 family structure
in this higher dimensional operator, we have to propose three copies
of $F_1,K_1,K_2,G_1,G_2$, like we have explained previously.
It can be noticed that the charge assignment and field content is such that,
while generating $\left(\frac{H^\dagger H}{M^2}\right)^2\bar{L}_LN_R\tilde{H}$,
the other higher dimensional term $\frac{H^\dagger H}{M^2}\bar{L}_LN_R\tilde{H}$
can not be possible in the low energy regime.

Now, it is easy to understand that the above described procedure can
be generalized to obtained some finite power of $\frac{H^\dagger H}{M^2}$
along with the $\bar{L}_LN_R\tilde{H}$. To generate the higher dimensional
term of Eq. (\ref{E:Dira}), we need the following eight different heavy fields:
complex scalar flavon fields ($F_i,i=1,\cdots,8$), singlet and doublet of
SU(2)$_L$ of the SM gauge group ($K_i,G_i,i=1,\cdots,8$).
Also, to get the full 3$\times$3 family structure in the $\bar{L}_LN_R\tilde{H}$,
we have to replicate the above heavy fields, with different charges, three times.
The number of heavy fields that we have described in the case of Dirac
neutrinos will be reduced in the case of Majorana neutrinos. To generate the
first term of Eq. (\ref{E:Majo}), we have to propose
four different heavy fields of scalar and fermionic type, which we have
described above. On top of this, to generate the second term of Eq. (\ref{E:Majo}),
we have to propose additional complex scalar field with a charge of $-2n$, where
$n$ is the U(1)$_F$ charge of the field $N_R$.

\section{Phenomenology}

The phenomenology of this model in the quark and charged lepton sector
is same as that described in \cite{babu_nandi}. In the neutrino
sector we get new phenomenological signals which will be described below.

First, from the UV completion of the higher dimensional operators of Eqs.
(\ref{E:Dira}) and (\ref{E:Majo}), we need some heavy weak-singlet and weak-doublet
ferminoic states. In the case of Dirac neutrinos, the number of these heavy
fermionic states is 8$\times$3 of both weak-singlet and weak-doublet fields.
Whereas in the Majorana case this number is 4$\times$3. The masses of these
heavy fermionic fields are in the range of 1$-$2 TeV. The heavy weak-singlet
$K$s have zero hypercharge, whereas the weak-doublet fields ($G$s) have hypercharge
$+1$. Hence, the heavy weak-doublet fields can be produced either through
$W$ or $Z$ boson fusion at the LHC. However, detection of weak-singlets is challenging
due to its sterile nature. These weak-singlet fermions can be produced in a collider
process through the decay of heavy weak-doublet fermions. One of the weak-singlets ($K_1$ in the previous
section) is bound to decay into an active neutrino and Higgs boson.
And similarly, one of the weak-doublets ($G_1$ or $G_2$ of the previous
section) is bound to decay into a right-handed neutrino and Higgs boson
state, if this is kinematically allowed. As a result of these interactions,
after the weak-singlet and weak-doublet fermions being produced in a collider,
they can ultimately cascade down to some number of neutrinos and Higgs bosons, depending on
the masses of these fermions.
In this model there are also heavy complex scalar fields,
which are SM gauge singlets and have masses of $\sim$1 TeV. The complex scalar flavon
fields ($F_i$), apart from their Yukawa couplings to heavy fermions, can
have quartic interactions with the Higgs doublet field. As a result
of this, each scalar flavon field can decay to a pair of Higgs bosons.
These complex scalar fields can be produced through the decay of
either weak-singlet or weak-doublet heavy fermion.

Apart from the phenomenology of heavy fermions and scalars in the
model, the local nature of the flavor group U(1)$_F$ give some more
phenomenology. Since the flavor group is gauged, there would be a gauge
boson $Z^\prime$ corresponding to the U(1)$_F$. The mass of $Z^\prime$ could
be in the TeV scale provided the gauge coupling ($g_F$) of U(1)$_F$ is ${\cal O}$(1).
In the UV completion of this model we have assumed that leptons are charged
under U(1)$_F$ and without any inconsistency we can assume that quarks are
singlets under U(1)$_F$. This particular charge assignment can relax constraints
on the gauge coupling $g_F$ due to Drell-Yan process. However, the
charge assignment of leptons can induce mixing between $Z$ and $Z^\prime$ at
a loop level. Since this mixing should be small \cite{pdg}, we may have to suppress
the coupling $g_F$. The details of these studies is beyond the limit of this work,
but we refer some recent works on the phenomenology of $Z^\prime$ \cite{Zprime}.
It is to be noticed that some of the phenomenology of $Z^\prime$ studied in \cite{LMN} can
also be applicable in this model. Finally, the heavy fermions are chiral
with respect to the flavor group and as a result there could be anomalies due to
gauged U(1)$_F$. To cancel these anomalies we have to propose some
additional fermions at the TeV scale through Green-Schwarz mechanism,
which is also suggested in \cite{LMN}.

Next, we focus on the phenomenology arising due to neutrino
masses of this model.
As said previously, one of the consequences of Majorana neutrinos is that
it generates neutrinoless double beta decay process at tree level. This
has been looked in various experiments,
see Refs. \cite{nu2beta},
for conducted and future proposed experiments.
The amplitude of this process depends on the quantity $m_{ee}=\sum_{i=1}^3
U_{ei}^2m_i$ \cite{vissani}, where $U=U_{\rm PMNS}$ and $m_i$ are the mass eigenvalues
of light neutrinos. The non-observation of this process has put an upper bound
on $m_{ee}$ to be $\sim$0.5 eV \cite{dob-beta}. In our particular case of
Majorana neutrinos, for $\epsilon=1/6.5$,
we have calculated the $m_{ee}$ of neutrinoless double beta decay process.
In the normal hierarchy, for $m_1=0$ and $m_1=0.5\times\sqrt{\Delta m^2_\odot}$,
$m_{ee}$ is $3.9\times 10^{-3}$ eV and $7.1\times 10^{-3}$ eV, respectively.
In the inverted hierarchy, for $m_3=0$ and $m_3=0.5\times\sqrt{\Delta m^2_{\rm atm}}$,
$m_{ee}$ is found out to be 0.048 eV and 0.054 eV, respectively. Note
here that we put Majorana phases to be zero in this calculation. An
interesting fact is that the values we get for the quantity $m_{ee}$ is
independent of right-handed neutrino masses, since they do not enter in
the construction of the PMNS matrix of this model.

Nonzero masses of neutrinos indicate oscillations in flavor neutrinos,
and hence we can have flavor changing processes such as $\mu\to e\gamma$. The
current upper bound on the branching ratio of this process is $2.4\times 10^{-12}$
at 90$\%$ C.L.\cite{mutogamma}.
For Dirac neutrinos, due to Glashow-Iliopoulos-Maiani (GIM) cancellation mechanism,
the amplitude for this process is highly suppressed, and the branching
ratio is well below the current upper limit. However, for Majorana neutrinos
the GIM cancellation mechanism is not valid, and one may wonder if we
can get appreciable branching ratio in this case. In the Majorana
case, we have Type-I seesaw mechanism and in this mechanism the branching ratio
for $\mu\to e\gamma$ has been derived in \cite{Ilakovac}. For a TeV scale
right-handed neutrino masses, we have found that the branching ratio
of this process is $\sim 10^{-31}$, which is significantly smaller than
the current upper limit. We believe that this small branching ratio in
our model is due to small admixture between light active and heavy
right-handed neutrinos.

Finally, we comment on possible restrictions on this model due to
$D^0-\bar{D}^0$ mixing. Since the Yukawa couplings are non-diagonal in
the quark sector, Higgs boson can generate flavor changing neutral
current processes. Among these the Higgs couplings to up and charm
quarks can give mass difference between $D^0-\bar{D}^0$ \cite{babu_nandi}.
The current upper limit on this mass difference is $2.35\times 10^{-14}$
at 2$\sigma$ level \cite{pdg}. In \cite{babu_nandi}, the Higgs
contribution to this mass difference is claimed to be $\approx 7\times 10^{-14}$,
which is now ruled out. However, this computation is done for a specific
values of $h_{12}^u=1.0$ and $h_{21}^u=0.5$ and also for a Higgs boson mass
of 200 GeV. Since now the Higgs boson has to be within 140 GeV 
\cite{ATLAS_new,CMS_new}, we have redone the computation with $h_{12}^u=1.06$, 
$h_{21}^u=0.5$ and for a Higgs mass of 130 GeV.
For $\epsilon=$ 1/6.5 and 0.16 we have
found $\Delta m_D^{\rm Higgs}=1.47\times 10^{-14}$ and $2.01\times 10^{-14}$,
respectively. For $\epsilon=0.17$ we have found that the mass difference
between $D^0-\bar{D}^0$ is $3.26\times 10^{-14}$ which exceeds the current
upper limit. However, $\epsilon=0.17$ can be made allowed by choosing different
set of $h_{12}^u$ and $h_{21}^u$ values. The price one may have to pay is adjusting
the Yukawa couplings to some decimal places. Here, it can be noticed that we have
fixed $h_{12}^u$ to two decimal places to get additional suppression compared
to that of \cite{babu_nandi}. The upper bound on $\epsilon$ we are getting
from $D^0-\bar{D}^0$ correlates with neutrino Yukawa couplings of this model. In Sec. 3 we
have quoted that in the inverted hierarchical case of Majorana neutrinos,
$\epsilon=0.17$ can give neutrino Yukawa couplings close to 1.0 rather than
for $\epsilon =$ 1/6.5 or 0.16. From the current data on $D^0-\bar{D}^0$ mixing,
the neutrino Yukawa couplings of this model are set to be on the higher side.
Further improvements on the $D^0-\bar{D}^0$ mixing can set limits on \cite{babu_nandi}
as well as neutrino sector of this model.

\section{Conclusions}

Some of the challenging problems in particle physics are the hierarchical
pattern of fermion masses and the difference in the mixing pattern for
quarks and leptons. A simple and elegant model \cite{babu_nandi} has
been proposed to explain the hierarchies in charged fermions and also the
mixing pattern in the quark sector. One of the parameters of this model is
$\epsilon=\frac{\langle H\rangle}{M}\sim 0.15$. After fixing this
parameter and for ${\cal O}(1)$ Yukawa couplings in the model, 
the charged fermion
masses and the CKM matrix have been obtained, in agreement with experimental
values.

We have generalized the model \cite{babu_nandi} to accommodate neutrino masses
and also obtain mixing pattern in lepton sector, i.e. PMNS matrix. We have
addressed both Dirac and Majorana masses for neutrinos in this model. To explain
the Dirac and Majorana masses consistently, we have proposed dimension 20 and 12
higher order terms, respectively. These higher order terms are shown to be arriving
from the UV completion of the model by having an additional flavor symmetry and
some TeV scale heavy fermions and scalars. After proposing higher order terms we
have done numerical analysis, where we have
shown that the atmospheric and solar neutrino mass scales and mixing pattern
in lepton sector can be consistently obtained for $\epsilon$ between 1/6.5 and 0.16.
From the $D^0-\bar{D}^0$ mixing, we have argued that we can set an upper bound on
$\epsilon$ to be around 0.17.

Since the $\epsilon$ parameter is tightly
constrained, the model has definite predictions for various collider processes.
From the neutrino sector, in the Majorana case, the effective mass
of the neutrinoless double beta decay, $m_{ee}$, has definite values
in this model. The values of $m_{ee}$, in the inverted neutrino mass hierarchical
case, are about an order less than the currently probed experimental values.
This case is interesting in the ongoing and future experiments
on the neutrinoless double beta decay process, to verify if the model
presented here is realistic. Apart from the signals in neutrino sector,
the UV completion of this model demands the presence of TeV scale weak-singlet
and weak-doublet fermionic as well as singlet scalar particles in the
theory. Detection of these particles is within reach of LHC experiment,
which can give smoking gun signals of this model.

\section*{Acknowledgments}
D.K.G. acknowledges partial support from the Department of Science and 
Technology, India under the grant SR/S2/HEP-12/2006.

\end{document}